\documentclass[aps,superscriptaddress,reprint,floatfix]{revtex4-1}

\usepackage[plainpages=false, colorlinks=true, anchorcolor=blue, linkcolor=blue, citecolor=blue, bookmarks=false]{hyperref}
\usepackage{amsmath, amsthm, amssymb}
\usepackage{graphicx}

\usepackage{booktabs}       
\usepackage{nicefrac}       
\usepackage{microtype}      
\usepackage{lipsum}
\usepackage{graphicx}
\usepackage{siunitx}
\usepackage{tikz}
\usepackage{float}
\usetikzlibrary{quotes,angles}
\usepackage[compat=1.1.0]{tikz-feynman}

\usepackage{soul}
\usepackage{xcolor}
\usepackage{dcolumn}
\usepackage{bm}
\usepackage[mathscr]{euscript}
\usepackage{mathrsfs}
\usepackage{amsbsy}

\begin{document}

\title{Axion detection through resonant photon-photon collisions}

\author{K. A. Beyer}
\email[Authors to whom correspondence should be addressed: ]{konstantin.beyer@physics.ox.ac.uk}
\affiliation{Department of Physics, University of Oxford, Parks Road, Oxford OX1 3PU, UK}
\author{G. Marocco}
\email{giacomo.marocco@physics.ox.ac.uk}
\affiliation{Department of Physics, University of Oxford, Parks Road, Oxford OX1 3PU, UK}
\author{R. Bingham}
\affiliation{Rutherford Appleton Laboratory, Chilton, Didcot OX11 0QX, UK}
\affiliation{Department of Physics, University of Strathclyde, Glasgow G4 0NG, UK}
\author{G. Gregori}
\affiliation{Department of Physics, University of Oxford, Parks Road, Oxford OX1 3PU, UK}
\newcommand{\gax}{g_{a\gamma\gamma}}
\newcommand{\gsc}{g_{s\gamma\gamma}}
\newcommand{\Gax}{\Gamma_{a\gamma\gamma}}

\begin{abstract}

We investigate the prospect of an alternative laboratory-based search for the coupling of axions and axion-like particles to photons. Here, the collision of two laser beams resonantly produces axions,
and a signal photon is detected after magnetic reconversion, as in light-shining-through-walls (LSW) experiments.
Conventional searches, such
as LSW or anomalous birefrigence measurements,
are most sensitive to axion masses for which substantial coherence can be achieved; this is usually well below optical energies. We find that using currently available high-power laser facilities, the bounds that can be achieved by our approach outperform traditional LSW at axion masses between $0.5-6$ eV, set by the optical laser frequencies and collision angle. These bounds can
be further improved through coherent scattering off laser substructures, probing axion-photon couplings down to $g_{a\gamma\gamma}\sim\SI{e-8}{GeV^{-1}}$, comparable with existing CAST bounds. Assuming a day long measurement per angular step, the QCD axion band can be reached.
\end{abstract}

\maketitle

\section{Introduction}

The Standard Model of particle physics allows for charge-parity (CP) violation in the electroweak and strong sectors. While the former is experimentally established, the latter is constrained by neutron electromagnetic dipole measurements to be negligibly small, with no natural explanation.
This is the strong CP problem.
One of the most well-motivated solutions to this problem is the axion proposal by Peccei and Quinn \cite{PhysRevLett.38.1440,PhysRevD.16.1791}. It has been subsequently noted that such models have a pseudo-Goldstone boson \cite{Weinberg:1977ma,Wilczek:1977pj}, the QCD axion, which can make up a substantial fraction or the entirety of the dark matter abundance  \cite{Preskill:1982cy,Abbott:1982af,Dine:1982ah}, and perhaps also be responsible for the 
initial magnetisation of the Universe
\cite{Miniati:2017kah}.
Similar pseudoscalars, known as axion-like particles (ALPs), readily arise in the low-energy spectrum of string theory \cite{Witten:1984dg,Arvanitaki:2009fg}. A number of experiments have already placed bounds \cite{Tanabashi:2018oca} on the available axion parameter space, with varying degrees of model-dependence; we use the term axion to refer to both the QCD axion and CP-conserving ALPs. 


An axion couples to Standard Model photons via
\begin{equation}
    \label{Eq:L_a}
    \mathcal{L}_p=-\frac{1}{4}F^{\mu\nu}F_{\mu\nu}+\frac{1}{2}\left(\partial_\mu a\right)\left(\partial^\mu a\right)-\frac{1}{2}m_a^2 a^2 +\frac{1}{4}g_{a\gamma\gamma}a \tilde{F}^{\mu\nu}F_{\mu\nu}
\end{equation}
with $a$ the axion field, $m_a$ the axion mass, $F$ the electromagnetic field-strength tensor and $\tilde{F}^{\mu\nu}=\frac{1}{2} \varepsilon^{\mu\nu\sigma\rho}F_{\sigma\rho}$ its dual. We will be interested in constraining the axion-photon coupling $g_{a\gamma\gamma}$. When interested in a CP-conserving scalar field $s$, the interaction term is instead
\begin{equation}
    \label{Eq:L_scalar}
    \mathcal{L}_s\supset \frac{1}{4}g_{s\gamma\gamma}s F^{\mu\nu}F_{\mu\nu}.
\end{equation}

The axion-photon coupling term gives rise to a cubic interaction of an axion with two photons, and can be expressed as $g_{a\gamma\gamma}a\mathbf{E}\cdot\mathbf{B}$ for pseudoscalars or $g_{a\gamma\gamma}a(E^2-B^2)$ for scalars. In the presence of an external magnetic field this leads to possibly observable axion-photon mass mixing  \cite{Sikivie:1983ip,Raffelt:1987im}.

We consider the mass range around $\SI{1}{eV}$, where the strongest bounds, excluding purely astrophysical arguments derived from stellar cooling times, are placed by the CAST experiment \cite{anastassopoulos2017new}, a helioscope sensitive to the axion flux produced by the Primakoff effect \cite{Pirmakoff:1951pj} in the Sun. Above \SI{1}{eV}, Primakoff coherent solar axion-to-photon conversion in crystals place stronger constraints \cite{bernabei2001search}. Due to the astrophysical origin of the axions and the consequential lack of control over the production, the possibility of model dependence must be seriously taken into account. The non-zero plasma frequency and high temperature conditions in the Sun affect the axion-photon effective coupling \cite{Jaeckel:2006xm}. Similar arguments apply to stellar cooling bounds. As such, we focus on complementary, model-independent bounds given by purely terrestrial experiments. 

Current searches like the OSQAR \cite{PhysRevD.92.092002} experiment implement a Sikivie-style LSW scheme \cite{Sikivie:1983ip} to detect the mass mixing. A laser beam propagating in a magnetic field can spontaneously create axions, which, being weakly coupled, pass through an intervening wall. Behind the wall, any axions entering the reconversion magnetic field may mix back into photons and be detected. 
This mass mixing is largest
if the magnetic field modes match the momentum exchanged between the axion and the photon. The large momentum exchange needed for conversion of non-relativistic axions suppresses the rate as the modes are set by the inverse length of the external magnetic field region.

Other laboratory axion searches include the PVLAS collaboration \cite{DellaValle:2015xxa}, looking at induced birefringence as a polarised laser beam traverses a magnetic field. This effect receives contributions from both QED \cite{Adler:1971wn} and axion \cite{Raffelt:1987im} interactions. The leading QED contribution is a one-loop process, which competes against the tree-level axion contribution. 
The exclusion bounds of PVLAS are thus limited by the requirement that the latter process is the dominant one.
Once axion interactions are smaller than the leading order Standard Model prediction, one must be careful of, for instance, hadronic contributions. Given the difficulty this presents in other precision measurements \cite{Melnikov:2019xkq}, disentangling each independent contribution becomes challenging. PVLAS experiments are close to this limit, suggesting new techniques are required to probe axion couplings below those bounds.

Our proposal aims to improve current laboratory searches by adopting an alternative axion-production mechanism. Here, we propose to generate axions via the collision of two laser beams. A similar idea involving photon-electron scattering has been explored in \cite{Dillon:2018ouq}, and light-by-light scattering at the LHC has placed bounds on heavy ALPs \cite{Knapen:2017ebd}. Lighter axions are better suited to experiments involving a large number of photons, rather than collider searches with large background signals of soft particles. Lasers are ideal for this pursuit, and we expect that the non-linear dependence on the number of photons will allow a better scaling compared to the linear dependence in traditional LSW experiments, as we will discuss further below. Further, resonant production at optical masses allows smaller couplings to be probed than in traditional LSW experiments. Using pulsed lasers allows for the rejection of most background by exploiting coincident timing. 
This minimal approach can achieve sensitivities comparable to PVLAS. With the implementation of spatial modes on the laser beam, coherent generation of axions can significantly improve our bounds on $g_{a\gamma\gamma}$ down to $\SI{e-8}{GeV^{-1}}$. These are comparable to bounds placed by CAST in the relevant mass region, but without any additional assumptions on the solar plasma.



\begin{figure}[h]
\includegraphics[]{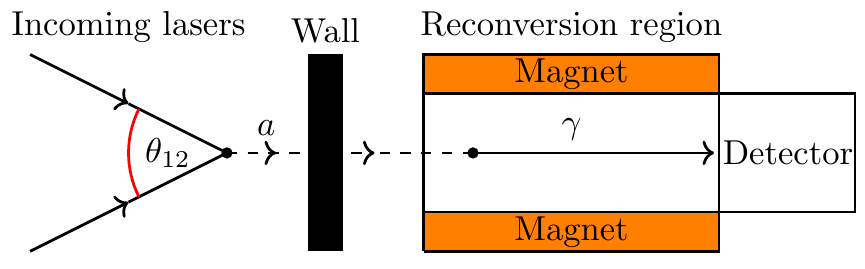}
\caption{A diagram of the experimental set-up. Photons from two lasers collide at in an interaction region, producing any hypothetical axions (either scalar or pseudoscalar). These pass through an intervening wall, and then reconvert into photons in the presence of a magnetic field. These photons are the signal detected.}
\label{fig:diagram}
\end{figure}

\section{Axion production and reconversion}

We consider two photons, each with energy $\omega_j$ ($j=1$, $2$), colliding at an angle $\theta_{12}$, as depicted in Fig.~\ref{fig:diagram}. The two incoming momenta define the scattering plane. The polarisation vectors can be expanded in terms of two basis vectors: one lying in the scattering plane and one perpendicular to it; both basis vectors are perpendicular to the momentum. The non-zero polarised matrix element for axion production is
\begin{equation}
    \begin{split}
    \label{Eq:MatrixElements}
    \mathcal{M}^p_{||,\perp}&=-2\gax\omega_1\omega_2\sin{\left(\frac{\theta_{12}}{2}\right)^2}, 
    \end{split}
\end{equation}
where a subscript $||$ refers to in-plane and $\perp$ to out-of-plane polarisation.

The corresponding cross-section is
\begin{equation}
\label{Eq:CrossSecCalc}
    \sigma=\frac{\pi}{4 \omega_1\omega_2(\omega_1+\omega_2)}\frac{\left|\mathcal{M}\right|^2}{\sqrt{2-2\cos{\theta_{12}}}}\delta\left( \omega_1+\omega_2-\omega_a \right)
\end{equation}
where $\omega_a=\sqrt{m_a^2+\omega_1^2+\omega_2^2+2\omega_1 \omega_2 \cos\theta_{12}}$ is the axion energy and only $\mathcal{M}=\mathcal{M}^p_{||,\perp}$ contributes. 

The appearance of a delta distribution is characteristic of energy-momentum conservation at a three point vertex. In a real experiment however the cross section is non-singular as the photons are provided by a laser beam of a finite pulse length $\tau$, and additionally are drawn from a Gaussian distribution
\begin{equation}
    \label{Eq:GaussianDist}
    f_j\left(\omega\right)=\frac{1}{\sqrt{2\pi\Delta_j^2}}\exp{\left[-\frac{\left(\omega-\omega^0_j\right)^2}{2\Delta_j^2}\right]}
\end{equation}
with central frequency $\omega^0_j$ and spectral width $\Delta_j \gtrsim \tau^{-1}$ dictated by the laser properties. Note also that in equation (\ref{Eq:CrossSecCalc}) we have implicitly taken the axion to be in an asymptotic state. This is a reasonable approximation given that its lifetime against stimulated decay to a photon is $\tau_{a\gamma\gamma}=64\pi /N_j \gax^2 m_a^3$ \cite{Bernard:1997kj} with $N_j$ photons in the interaction region, which is always much longer than the laser pulse-length, $\tau \simeq \SI{230}{eV^{-1}}$. We thus treat the axion resonance as a delta distribution to be integrated against the Gaussian (\ref{Eq:GaussianDist}). 

We now define the scattering probability as
\begin{equation}
    \label{Eq:Prob}
    P_{\gamma\gamma\rightarrow p}=\frac{1}{V}{\sqrt{2-2\cos{\theta_{12}}}}\int_0^{\tau}\sigma dt,
\end{equation}
where $V$ is the interaction volume of the two lasers. The angular dependence describes the flux of one beam through the other. Given the probability for pseudoscalar production (\ref{Eq:Prob}), the total number of incoherently produced axions is 
\begin{equation}
    \label{Eq:AxNumb}
    N_p=N_1 N_2 P_{\gamma\gamma\rightarrow p},
\end{equation}
where $N_j$ is the total number of photons of beam $j$ in the interaction region.


Our proposed scheme for axion production can also be applied to scalars with a coupling given by equation (\ref{Eq:L_scalar}). The associated non-zero matrix elements are
\begin{align}
    \label{Eq:ScalarSMatrix}
    \mathcal{M}^s_{\perp,\perp} &=\mathcal{M}^s_{||,||}/ \cos{\theta_{12}}=-2g_{s\gamma\gamma}\omega_1\omega_2\sin{\left(\frac{\theta_{12}}{2}\right)}^2.
\end{align}
Note that scalars and pseudoscalar 
couple with opposite photon polarizations, making it possible
to separate the two signatures independently. This is challenging to realise with other laser-based searches.
The photon to scalar conversion probability is completely analogous to equation (\ref{Eq:Prob}), with $p\rightarrow s$.

\begin{figure}[h]
\centering
\includegraphics[]{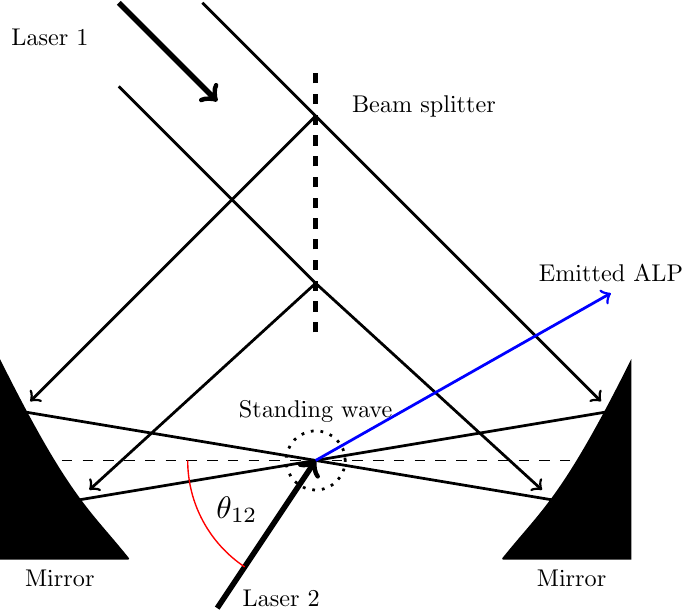}
\caption{A standing wave is formed in the circled region by passing a single laser pulse through a 50-50 beam splitter, each part of which then reflects of an off-axis parabolic mirror, as outlined in \cite{Heinzl:2006xc}.}
\label{fig:standingwave}
\end{figure}


The  probability of generating axions can be further enhanced by implementing spatial structure in one of the beams. In our proposed scheme, coherence must be achieved for large momentum transfer $q$, since axions become more non-relativistic as their mass approaches the photon frequency. In this case, a traditional LSW experiment suffers from interference effects which destroy coherence along the length $\ell$ of the interaction region when $q \ell \gtrsim 1$. This is avoided for the interaction of the probe beam with a standing wave, which can lead to a scattering amplitude somewhat similar to Bragg diffraction off of a light grating \cite{martin1988bragg}. In this case, a probe laser interacts with a standing wave produced by the other beam as shown in Fig. \ref{fig:standingwave} (see also Ref. \cite{Heinzl:2006xc}). The standing wave is realised by reflecting a split laser beam off two off-axis mirrors such that they are counter-propagating at  a common focus. Scattering is then coherent within any single half-period of the standing wave, leading to an overall cubic scaling with photon number (see below).

Coherence effects have been used in the conversion of solar axions through Bragg diffraction with the electric field of a crystal \cite{paschos1994proposal,bernabei2001search}, but, as the enhancement relies on the solar axions being converted into X-rays, it is not compatible with axions produced by optical lasers considered here. Additionally, in our case the momentum transfer and the spatial structure along the direction of beam propagation are intrinsically linked since they are both given by the frequency of the light in the standing wave, precluding coherence along the whole laser beam without some kind of external temporal modulation.


Let us consider a probe photon beam interacting with a standing wave (Fig. \ref{fig:standingwave}). We approximate the probe beam photons as point-like, since the probe 
wavelength is much smaller than the standing wave wavelength. The structure factor associated with the momentum-transfer ($\mathbf{q}$) is then given by
\begin{equation}
I(\mathbf{q}) = \sum_{i,j=1}^{N_1} e^{i\mathbf{q}\cdot (\mathbf{x}_i-\mathbf{x}_j)},
\end{equation}
where $\mathbf{x}_i$ are the positions of the standing wave photons in the interaction region. The momentum transfer $q$ is simply the momentum $\pm k$ of a photon in beam 1 for our elastic scattering. Generically, the sum over these phases will be incoherent, and so only the diagonal piece of the matrix will contribute, leading to a scaling proportional to $N_1$. To achieve some form of coherence, the off-diagonal pieces ought to constructively interfere, providing a quadratic scaling. We have
\begin{align}
   I(\mathbf{q})=\left(\frac{2N_1}{L}\right)^2\left|\int_{-\frac{L}{2}}^{\frac{L}{2}}dx \cos^2(kx+\phi)e^{-ikx}e^{-\frac{x^2}{2w^2}}\right|^2,
\end{align}
where $L$ is the width of the probe beam and the Gaussian models the intensity fall-off of the standing wave, governed by its Rayleigh length $w$. The phase $\phi$ that the probe photon sees is dependent on where the probe beam falls on the standing wave. To achieve the largest coherence, we must achieve a focus such that $L \simeq \lambda_1/2$. For such a scenario, we have $w \gg L$, and we find that the structure factor is rather insensitive to $\phi$, varying by a factor of 4 over its range.
Once created the axions must be reconverted into photons to be detected. Since the axions are on-shell, light, and weakly coupled, they propagate over a macroscopic distance. 
Such a decay into a visible photon can in principle be enhanced by stimulated processes, similar to what has been proposed for QED precision tests with three colliding lasers \cite{Lundstrom:2005za}. An additional (third) beam is introduced into the interaction region and can induce a stimulated decay into a photon. However, given available photon numbers, stimulated decay is negligible for any interesting coupling constant \cite{Bernard:1997kj}. 
Hence, we will adopt the usual approach of reconverting the axions in a magnetic field placed after an intervening wall.


As for any LSW experiment, the reconversion region is filled by a background magnetic field $B$ over a length $L$. The mixing probability is given by \cite{Sikivie:1983ip,Adler:2008gk}
\begin{equation}
    \label{Eq:RecProb}
    P_{a\rightarrow\gamma}=\frac{\gax^2 B^2 L^2}{\beta_a}\left(\frac{\sin{qL/2}}{qL}\right)^2,
\end{equation}
with the momentum transfer $q = \sqrt{\omega_a^2-m_a^2}-\omega_a$, the axion energy $\omega_a = \omega_1+\omega_2$, and $\beta_a$ the axion velocity. The sinc function describes the momentum modes present in the magnetic field, arising from taking the spatial Fourier transform of the magnetic field, assumed constant over the length $L$, and zero otherwise.
Adler's photon splitting theorem, which would cause the amplitude to vanish for $L \rightarrow \infty$ \cite{adler1971photon}, is thus avoided as the magnetic field can transfer momentum.


If the total momentum of the incoming lasers points in the same direction, then the reconversion region can be left unchanged from shot to shot. Due to the spectral width of the incoming beams the cross section of the magnetic region has to be sufficiently large to allow detection of any axion in the emission cone. Finally, to avoid edge effects, the length of the magnetic region should be at least as long as its width.

The number of detected photons is finally given by 
\begin{equation}
    \label{Eq:Signal}
    N_\gamma= \int I(q)N_2 P_{\gamma\gamma\rightarrow a} P_{a\rightarrow\gamma} f_1\left(\omega_1\right)f_2\left(\omega_2\right)d\omega_1 d\omega_2,
\end{equation}
where we have assumed the laser beam frequency distribution is given by equation (\ref{Eq:GaussianDist}),
and $\omega_1$ and $\omega_2$ are the standing wave and probe frequencies, respectively.

\section{\label{Sec:Signal}Projected bounds}

\begin{figure}[hb]
    \includegraphics[width=0.47\textwidth]{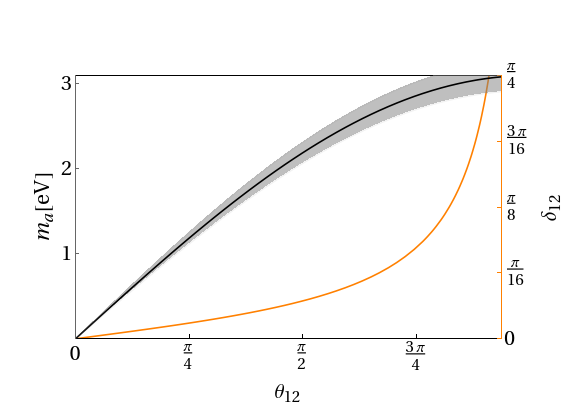}
    \caption{The orange curve plots the angular step size $\delta_{12}$ against the chosen angle $\theta_{12}$ for each shot. Here the spectral width is $10/\tau$. On the left axis, the black curve indicates the central mass probed for a given $\theta_{12}$ and $\omega_1^0 = \omega_2^0=  \SI{1.55}{eV}$, the shaded region indicates the width $\pm \delta m_a$. Assuming a minimum possible step-size $\delta_{12}\gtrsim \SI{1}{\degree}$, the full mass range can be scanned in $\sim 30$ shots. This step size imposes a lower bound on $\theta_{12}\gtrsim \SI{0.4}{rad}$ corresponding to $m_a \gtrsim \SI{0.6}{eV}$. 
    }
    \label{fig:Angles}
\end{figure}

The proposed experiment is most sensitive to a hypothetical axion whose mass matches the central frequencies of the two beams, that is, $m_a=\sqrt{2\omega_1\omega_2(1-\cos{\theta_{12}})}$, by 4-momentum conservation. Due to the non-zero bandwidth of the incoming laser beams, axions of different masses can be produced with decreasing
probability moving away from the central frequencies of the beams. The width of the sensitivity region, which we define as the mass region within which the coupling bound varies by less than $\sqrt{2}$, for the case of two beams of similar spectral distribution, is $\delta m_a/m_a \simeq \Delta_1/\omega_1+\Delta_2/\omega_2$.
Hence in order to scan across a range of possible axion masses, we must take shots in increments of angle $\delta_{12}$ given by
\begin{equation}
    \label{Eq:Angles}
    \delta_{12} \simeq 2\frac{\delta m_a}{m_a}\left(\csc{\theta_{12}}-\cot{\theta_{12}}\right).
\end{equation}
This angular step size and corresponding mass region is shown in Figure \ref{fig:Angles}. 
We require that the spread in momentum is small enough so that an axion will enter the finite aperture of the reconversion region. For the values assumed in the paper, the aperture must be $\phi\geq\SI{0.03}{rad}$ which would require a magnetic field width $\leq\SI{30}{cm}$ at a realistic detector distance $\leq\SI{10}{m}$.


We now estimate the experimental feasibility of the proposed setup by calculating the expected number of photons produced. We consider a transition-edge sensor (TES) similar to the ALPS II experiment \cite{bastidon2016quantum,spector2016alps} capable of single photon detection. The effective exposure time of our experiment is set by the readout time of the detector, which is around $\sim\SI{1}{\mu s}$. The dark count rate of such detectors over the effective exposure time is $10^{-10}$ and efficiencies are high enough to detect single photons. The dominant background will come from blackbody radiation which can easily be rejected via their energy range, which lies outside the axion energy range, or coincident timing with the laser beam.

We see from equation \ref{Eq:Signal} that to
increase the signal we
must maximise the laser intensity. Some of the current or near-commissioning high-power laser facilities can achieve numbers of up to $\sim 10^{21}$ photons per beam. The Aton 4 ($\SI{10}{PW}$) laser at ELI-beamlines, for example, can operate at a central optical frequency of $\SI{1.55}{eV}$ with spectral width of $\Delta= \SI{0.05}{eV}$, and pulse length of $\tau = \SI{100}{fs}$. We consider the scenario in which both beams are focused down to their respective wavelengths. Frequency doubling and tripling is also possible, as is the beam splitting in two independent arms. We consider the beams to be focused down to their respective photon wavelengths. Finally, we consider, in the reconversion region, a magnetic field of strength $\SI{10}{T}$, although the limits on $g_{a\gamma\gamma}$ scale only as $B^{-1/2}$. Due to the pulsed nature of the axion conversion, the magnetic field can be pulsed as well. Such high fields require superconducting magnets. The design of laser target chambers sets a lower bound to the distance between the interaction region and the detector of a few meters. To ensure an aperture of at least $\SI{0.03}{rad}$ we assume a magnetic field of $\SI{30}{cm}$ in each dimension to evaluate equation \ref{Eq:Signal}. Note that the length does not play a significant role as long as it is larger than the other spatial dimensions as when $qL\gg1$, the $\sin^2(qL/2)$ in (\ref{Eq:RecProb}) oscillates quickly and the laser spectral width effectively averages over the oscillations. In this case the length drops out of the probability.
\begin{figure}[ht]
    \includegraphics[width=0.47\textwidth]{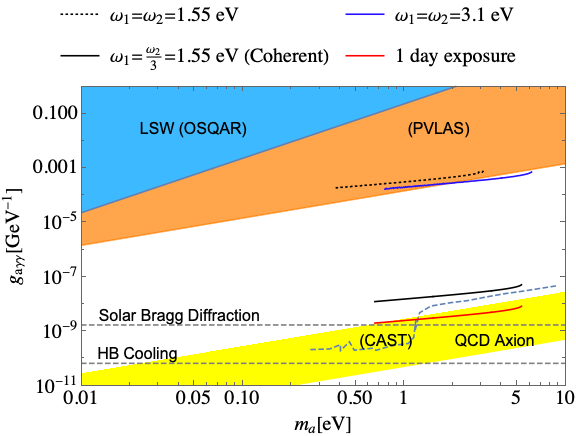}
    \caption{Exclusion plot for axion parameter space. The light blue region shows existing bounds from the OSQAR experiment \cite{Ballou:2015cka}; the orange region is excluded by PVLAS \cite{DellaValle:2015xxa}; the dashed blue line depicts CAST constraints \cite{anastassopoulos2017new}; the lower horizontal dashed line comes form stellar cooling lifetimes \cite{ayala2014revisiting} and the upper from solar Bragg diffraction experiments \cite{bernabei2001search}. The dotted black and dark blue lines correspond to our proposal performed at $\omega=\SI{1.55}{eV}$ and the first harmonic, respectively, with ELI parameters and no substructure coherence. The black line shows the possible bounds using standing wave coherence and the red line indicates the same parameters but $1$ day of shots per angular step instead of a single shot. The region above the line is excluded in each case. The QCD axion region indicates particular theoretical predictions for where the axion might be, given dark matter abundances \cite{di2017redefining}.}
    \label{fig:ExclPlot}
\end{figure}

Figure \ref{fig:ExclPlot} shows the axion exclusion limits we expect to achieve with the proposed experiment. The bounds are calculated assuming a single shot for each appropriately spaced angular step size. The whole experiment entails around $30$ shots. We see that our technique is able to access regions of parameter space that are not already excluded by existing LSW experiments. 
The lines in Figure \ref{fig:ExclPlot} must be interpreted as the upper bound on $g_{a\gamma\gamma}$ and, for given laser parameters, they are constructed by changing the scattering angle as discussed earlier so that we constrain an order of magnitude in $m_a$. 
The scalar exclusion plot gives very similar bounds.
The two upper lines in Figure \ref{fig:ExclPlot} are obtained without any coherent enhancement. While PVLAS provides slightly better bounds, however, as discussed above, these suffer from irreducible Standard Model background limitations. Our proposed searches, on the other hand, have no such background.
When scattering from coherent laser substructures is implemented, our bounds on the axion-photon coupling are competitive with CAST in the $1-5$ eV mass range, and model independent. Astrophysical bounds, as is the case for all laboratory experiments in this mass range, provide more stringent bounds, again with the caveat that they are model dependent.

Despite the unfavourable scaling of the bounds with the number of shots, the red curve shows that the QCD band can be reached comfortably with $1$ day run time per step size.


Considering the progress in laser technology over the past decades \cite{strickland1985compression}, it is not inconceivable that laser intensities would grow further over the coming years to allow new parameter space to be probed. Given the cubic scaling with photon number for coherent scattering, these advances can achieve a significant improvement of the bounds presented here, and more so compared to other LSW experiment.
The mass range can also be extended by using XFELs, for which a higher degree of frequency tuning is possible. This would allow for easier scanning of parameter space.


The experimental parameters we consider here do not maximise the coherence factor. In fact, the reconversion efficiency can be improved if the magnetic field were spatially varying, such that its Fourier transform components match the momentum transfer in the axion-photon transition. Moreover, the refractive index of the reconversion chamber could also be varied to reduce the difference in the axion mass and the effective photon mass, which would decrease the required momentum transfer. We have not yet included these effects in our analysis, and as such our results can be considered a lower bound on the performance of this technique at high-intensity laser facilities. 

\begin{acknowledgments}
We would like to thank Prof Subir Sarkar, Rami Shelbaya and especially Charlotte Palmer for many useful discussions. We thank Prof Willy Fishler for comments on an early version of the paper.
The research leading to these results has received funding from AWE plc. British Crown Copyright 2019/AWE.
\end{acknowledgments}

\bibliography{references}  


\end{document}